\documentclass[
letterpaper,
reprint,           
twocolumn,
superscriptaddress,
amsmath,           
amssymb,           
aps,               
prd,               
notitlepage,       
longbibliography,  
floatfix,          
nofootinbib,
]{revtex4-1}

\usepackage{subfigure}
\usepackage{amsmath}
\usepackage{lipsum}
\allowdisplaybreaks[4]
\usepackage{cancel}
\usepackage{extarrows}
\usepackage{tensor}     
\usepackage{float}
\usepackage[final]{graphicx}   
\usepackage[
colorlinks=true,        
citecolor=blue,         
linkcolor=blue,         
urlcolor=blue           
]{hyperref}             
\usepackage{bm}         
\usepackage{xcolor}     
\usepackage{lipsum}
\usepackage{color}      
\usepackage[utf8]{inputenc} 
\usepackage[section]{placeins} 
\usepackage{appendix}
\usepackage{units}
\usepackage[capitalise]{cleveref}
\usepackage{units}
\usepackage{url}
\newcommand{\nc}{\newcommand*}

\nc{\xbar}{\bar{x}}
\nc{\rhoeq}{\rho_{\mathrm{eq}}}
\nc{\zeq}{z_{\mathrm{eq}}}
\nc{\tla}{\tilde{\lambda}}
\nc{\bt}{\beta}
\nc{\dt}{\delta}
\nc{\Dt}{\Delta}
\nc{\vj}{\vec{j}}
\nc{\vl}{\vec{l}}
\nc{\hx}{\hat{x}}
\nc{\hy}{\hat{y}}
\nc{\bj}{\bm{j}}
\nc{\mJ}{\mathcal{J}}
\nc{\mP}{\mathcal{P}}
\nc{\Msun}{M_\odot}
\nc{\app}{\approx}
\nc{\av}[1]{\langle #1 \rangle}
\nc{\eq}[1]{Eq.~\eqref{#1}}
\nc{\al}{\alpha}
\nc{\Xstar}{X_{\ast}}
\nc{\fpbh}{f_{\mathrm{pbh}}}
\nc{\vth}{\vec{\theta}}
\nc{\vla}{\vec{\lambda}}
\nc{\vd}{\vec{d}}
\nc{\Mmin}{M_{\mathrm{min}}}
\nc{\rmd}{\mathrm{d}}
\nc{\mmin}{{m_{\mathrm{min}}}}
\nc{\mmax}{{m_{\mathrm{max}}}}
\nc{\mR}{\mathcal{R}}
\nc{\tmR}{\tilde{\mathcal{R}}}
\nc{\s}{\sigma}
\nc{\ogw}{\Omega_{\mathrm{GW}}}
\nc{\addref}{[\textcolor{red}{add ref}] }
\nc{\Om}{\Omega}
\nc{\gm}{\gamma}
\nc{\Gm}{\Gamma}
\nc{\gpcyr}{\mathrm{Gpc}^{-3}\,\mathrm{yr}^{-1}}
\nc{\Eq}[1]{Eq.~\eqref{#1}}
\nc{\Fig}[1]{Fig.~\ref{#1}}
\nc{\Table}[1]{Table~\ref{#1}}
\nc{\lvc}{LIGO/Virgo} 
\nc{\Sec}[1]{Sec.~\ref{#1}}
\nc{\eg}{\textit{e.g.~}}
\nc{\SNR}{\mathrm{SNR}}
\nc{\be}{\mathbf{\epsilon}}
\nc{\bn}{\mathbf{n}}
\nc{\bd}{\mathbf{d}}
\nc{\ba}{\mathbf{a}}
\nc{\eps}{\epsilon}
\nc{\bnu}{\mathbf{\nu}}
\nc{\mb}{\mathbf}
\nc{\bbt}{\mathbf{t}}
\nc{\bth}{\mathbf{\theta}}
\nc{\bep}{\mathbf{\epsilon}}
\nc{\uni}{\mathrm{U}}
\nc{\logu}{\operatorname{\mathrm{log-U}}}
\nc{\RN}{\mathrm{RN}}
\nc{\BN}{\mathrm{BN}}
\nc{\GN}{\mathrm{GN}}
\nc{\mcN}{\mathcal{N}}
\nc{\GWB}{\mathrm{GW}}
\nc{\yr}{\mathrm{yr}}
\nc{\Am}{\mathcal{A}}
\nc{\Dm}{\mathcal{D}}
\nc{\Hm}{\mathcal{H}}
\nc{\mc}{\mathcal{M}}
\nc{\sovast}{Soviet Ast.}

\nc{\mrm}{\mathrm}
\nc{\BE}{B\scriptsize{AYES}\normalsize{E}\scriptsize{PHEM}\normalsize  }

\nc{\Ostgw}{\Omega_{\mathrm{GW}}^{\mathrm{ST}}}
\nc{\Ottgw}{\Omega_{\mathrm{GW}}^{\mathrm{TT}}}
\nc{\Ovlgw}{\Omega_{\mathrm{GW}}^{\mathrm{VL}}}
\nc{\Oslgw}{\Omega_{\mathrm{GW}}^{\mathrm{SL}}}
\nc{\cosxi}{\beta}

\nc{\gmPL}{\gamma_{\mathrm{PL}}}
\nc{\APL}{A_{\mathrm{PL}}}

\def\({\left(}
\def\){\right)}
\def\[{\left[}
\def\]{\right]}

\def\e{\begin{equation}}
\def\q{\end{equation}}
\def\m{\begin{eqnarray}}
\def\n{\end{eqnarray}}
\nc{\red}[1]{\textcolor{red}{#1}}

\begin{document}


\title{Impacts of Gravitational-Wave Background from Supermassive Black Hole Binaries on the Detection of Compact Binaries by LISA}

\author{Fan Huang}
\email{huangfan@itp.ac.cn}
\affiliation{CAS Key Laboratory of Theoretical Physics, 
    Institute of Theoretical Physics, Chinese Academy of Sciences,Beijing 100190, China}
\affiliation{School of Physical Sciences, 
    University of Chinese Academy of Sciences, 
    No. 19A Yuquan Road, Beijing 100049, China}
\author{Yan-Chen Bi}
\email{biyanchen@itp.ac.cn}
\affiliation{CAS Key Laboratory of Theoretical Physics, 
    Institute of Theoretical Physics, Chinese Academy of Sciences,Beijing 100190, China}
\affiliation{School of Physical Sciences, 
    University of Chinese Academy of Sciences, 
    No. 19A Yuquan Road, Beijing 100049, China}
\author{Zhoujian Cao}
\email{zjcao@amt.ac.cn}
\affiliation{Institute of Applied Mathematics, Academy of Mathematics and Systems Science, Chinese Academy of Sciences, Beijing 100190, China}
\affiliation{School of Fundamental Physics and Mathematical Sciences, Hangzhou Institute for Advanced Study, UCAS, Hangzhou 310024, China}
\author{Qing-Guo Huang}
\email{huangqg@itp.ac.cn}
\affiliation{CAS Key Laboratory of Theoretical Physics, 
    Institute of Theoretical Physics, Chinese Academy of Sciences,Beijing 100190, China}
\affiliation{School of Physical Sciences, 
    University of Chinese Academy of Sciences, 
    No. 19A Yuquan Road, Beijing 100049, China}
\affiliation{School of Fundamental Physics and Mathematical Sciences, Hangzhou Institute for Advanced Study, UCAS, Hangzhou 310024, China}


\begin{abstract}
In the frequency band of Laser Interferometer Space Antenna (LISA), extensive research has been conducted on the impact of foreground confusion noise generated by galactic binaries within the Milky Way galaxy. Additionally, the recent evidence for a stochastic signal, announced by the NANOGrav, EPTA, PPTA, CPTA and InPTA, indicates that the stochastic gravitational-wave background generated by supermassive black hole binaries (SMBHBs) can contribute a strong background noise within in LISA band. Given the presence of such strong noise, it is expected to have a considerable impacts on LISA's scientific missions. In this work, we investigate the impacts of the SGWB generated by SMBHBs on the detection of massive black hole binaries (MBHBs), verified galactic binaries (VGBs) and extreme mass ratio inspirals (EMRIs) in the context of LISA, and find it crucial to resolve and eliminate the exceed noise from the SGWB to ensure the success of LISA's missions.

\end{abstract}
\maketitle

\section{Introduction} 

Laser Interferometer Space Antenna (LISA) is the space-borne gravitational wave (GW) detector operating in the frequency band of approximately $10^{-4} \sim 10^{-1}$  Hz \cite{Armano:2016bkm,LISA:2017pwj,LISA:2022yao}. This low-frequency band is abundant in a variety of GW sources that will enable us to observe the universe in a new and unique way, yielding nova insights in a wide range of topics in astrophysics and cosmology.

LISA has proposed a multitude of scientific objectives (SOs) associated with the necessary observation requirements for their fulfillment. Those observation requirements are in turn related to mission requirements (MRs) pertaining to noise performance, mission duration \textit{etc}, which requires calculation of signal-to-noise-ratio (SNR) for assessment \cite{LISA:2017pwj}. Different noise performance levels can lead to significant variations in SNR for a specific GW source. Meanwhile, the detectability and parameter measurement accuracy of this source will also be effected by the noise.

According to the 2017 LISA design \cite{LISA:2017pwj}, the strain sensitivity curves of LISA is the combination of predicted Michelson-equivalent sensitivity and stochastic gravitational wave background (SGWB) noise. In previous papers,
the foreground noise around $10^{-3}$ Hz due to galactic binary \cite{Barack:2004Sn,Robson:2018Sn,Cornish:2017vip} and the SGWB above $10^{-3}$ Hz from stellar origin black holes (SOBHs) \cite{Chen:2018rzo} were discussed. And the recent evidence for stochastic signal consistent with SGBW in the spectrum from $10^{-9}\sim10^{-1}$ Hz, announced by North American Nanohertz Observatory for Gravitational Waves (NANOGrav) \cite{NANOGrav:2023gor,NANOGrav:2023hde,NANOGrav:2023hvm}, the European Pulsar Timing Array (EPTA) in collaboration with the Indian Pulsar Timing Array (InPTA) \cite{EPTA:2023fyk,EPTA:2023sfo}, the Parkes Pulsar Timing Array (PPTA) \cite{Reardon:2023gzh,Zic:2023gta} and the Chinese Pulsar Timing Array (CPTA) \cite{Nan:2011um,Xu:2023wog}, indicates that the SGWB due to Supermassive black hole binaries (SMBHBs) can significantly contribute a background noise in LISA frequency band \cite{NANOGrav:2023hfp,EPTA:2023xxk,Bi:2023tib,Ellis:2023oxs}, bringing potential challenges to the LISA mission. However, this influence has not been investigated well before.

In this study, we utilize the up-to-date SGWB data from SMBHB indicated by NANOGrav 15-year data set, to analysis its impacts on the LISA mission. This paper is organized as follows: We firstly introduce the sensitivity of LISA and present the analytic-fit sensitivity curve we adopted in Section \ref{LISA}. Since the majority of individual LISA sources will be binary systems covering broad range of masses \cite{LISA:2017pwj,LISA:2022yao}, we then address the impacts from the SGWB generated by SMBHBs on the detection of compact binaries in the LISA mission. More specifically, we examine the effect on the detection of massive black hole binaries, verified galactic binaries (VGBs) and extreme mass ratio inspirals (EMRIs) in Sections \ref{MBHBs} and \ref{VGBandEMRI}, respectively. Finally, we draw our conclusion in Section \ref{Conclusion}.

\section{Sensitivity of LISA} \label{LISA} 

Here we adopt the analytic-fit sensitivity curve $S_a(f)$ for Michelson-style LISA data channel given by \cite{Robson:2018Sn}, as follows
\begin{widetext}
\e
S_a (f)= 
\frac{10}{3L^2}\( P_{\rm OMS}(f)+2(1+\cos^2(f/f_*))\frac{P_{\rm  acc}(f)}{(2\pi f)^4 } \) \left( 1+\frac{6}{10} \left( \frac{f}{f_*} \right) ^2 \right),
\q
\end{widetext}
where the transfer frequency $f_*=19.09$ mHz and arm length $L=2.5 \times 10^6$ km. In addition to the instrument noise, the galactic confusion noise, also called stochastic foreground, generated by unresolved galactic binaries will contribute an extra noise $S_c(f)$ to sensitivity, yielding the LISA's effective strain sensitivity curve $S_n(f)$ becomes the sum of $S_a(f)$ and $S_c(f)$ \cite{LISA:2017pwj,Barack:2004Sn,Robson:2018Sn,Cornish:2017vip}. The detailed expressions of single-link metrology noise $P_{\rm OMS}(f)$ and single test mass acceleration noise $P_{\rm  acc}(f)$, and the galactic confusion noise $S_c(f)$ are given in \cite{Robson:2018Sn}. Such additive of strain sensitivity indicate that we could calculate the impacts of SWGB via adding up its corresponding strain sensitivity $S_{\rm GW}(f)$ to $S_n(f)$.

Following methods proposed by \citep{Barack:2004Sn,Robson:2018Sn}, we define the noise strain sensitivity due to the SGWB as 
\e
S_{\rm GW}(f) = \frac{3H_0^2}{2\pi^2} \frac{\ogw(f)}{f^3} ,
\q
which can be added to the strain sensitivity of LISA with galactic confusion noise $S_n(f)$ to obtain an effective strain sensitivity $S_{\rm eff}(f) = S_n(f) + S_{\rm GW}(f)$ \cite{LISA:2017pwj,Barack:2004Sn,Robson:2018Sn,Cornish:2017vip}, and then the SNR effected by the present of GW background can be written as
\e
\SNR = 2 \[ \int df \frac{\vert \tilde{h}(f) \vert^2}{S_{\rm eff}(f)}  \]^{1/2},
\label{snr}
\q
where $\tilde{h}(f)$ is the frequency domain representation of time-domain waveform $h(t)$, which encodes exquisite information of intrinsic parameters of GW source.

\begin{figure}[htbp]
	\centering
	\includegraphics[width=0.5\textwidth]{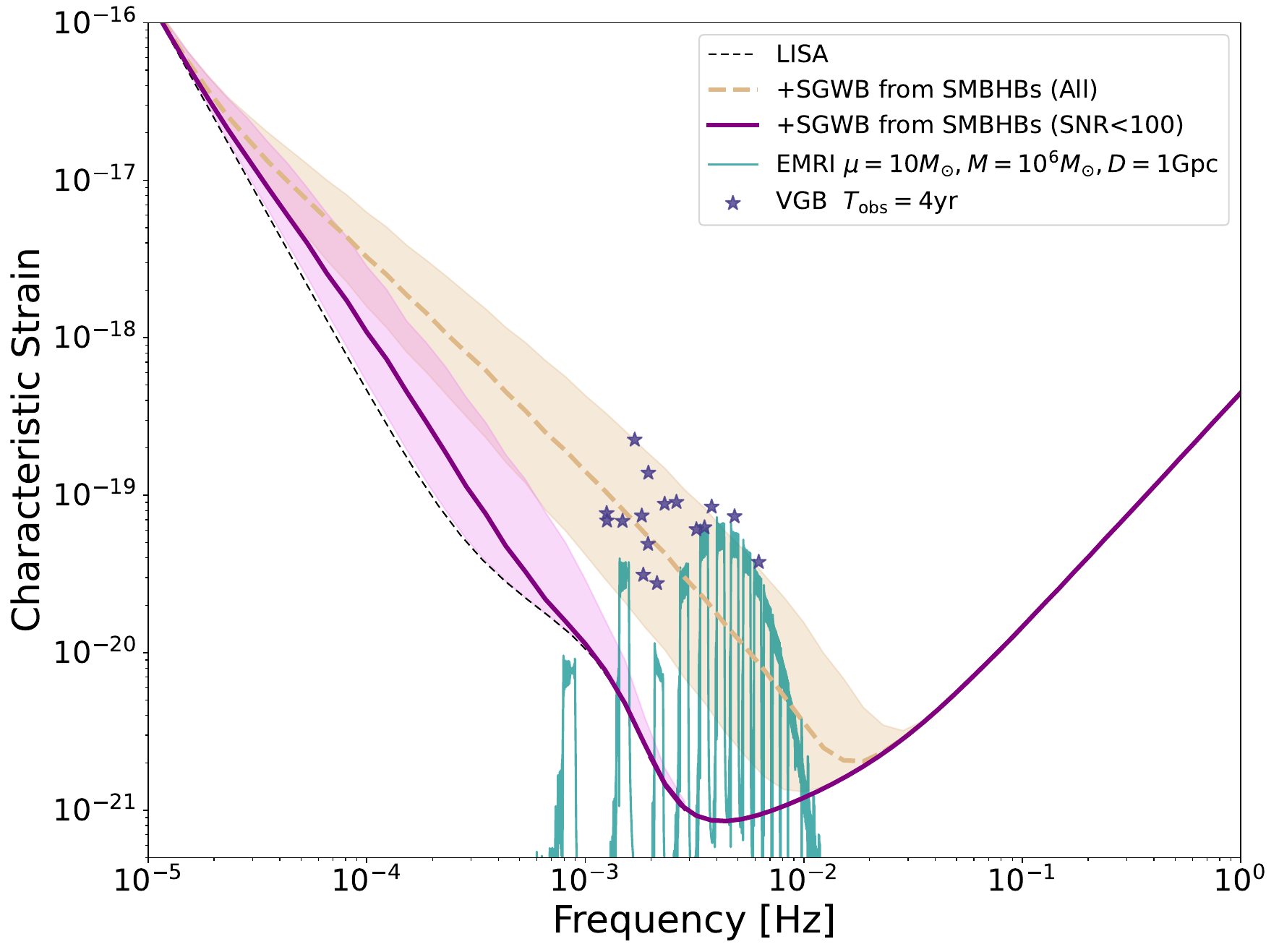}
	\caption{\label{charac_strain} 
	The expected effective characteristic strain $S_{\rm eff}(f)$ in the frequency range $[10^{-5},1]$ Hz. The burlywood line represents the effective characteristic strain derived from the SGWB of all SMBHBs, while the purple line illustrates the effective characteristic strain obtained from the SGWB of SMBHBs after excluding those with SNR $\ge$ 100. The shaded regions in both cases indicate the $90\%$ credible intervals. Additionally, we depict the expected characteristic strain curves of the LISA with black dashed lines. The darkcyan line corresponds to the characteristic strain produced by an illustrative EMRI signal \citep{Chua:2017EMRI}, while the star marker represents VGB signals \citep{Kupfer:2023nqx, VGBdata}.}
\end{figure}

In \Fig{charac_strain}, we present the effective characteristic strain $S_{\rm eff}(f)$ influenced by the SGWB originating from SMBHBs together with the specific GW signals of VGBs and an illustrative example of EMRIs. The effective characteristic strain, which represents the cumulative impact of all SMBHBs, is depicted as the burlywood-colored region in \Fig{charac_strain}. Notably, within the frequency range below several $10^{-2}$ Hz, it experiences a dramatic increase, partially obscuring specific signals associated with VGBs and EMRIs. It is believed that GW signals from SMBHBs with SNR equal to or greater than $\mathcal{O}(100)$ (Here we set 100 as criteria to be on the safe side) can be resolved and eliminated from the SGWB, as indicated in \citep{LISA:2017pwj,Pitte:2023ltw}. Our approach involves using $S_{\rm eff}(f)$, which accounts for contributions from all SMBHBs, as a baseline. We iteratively eliminate the contributions from SMBHB events with SNR $\ge 100$ and derive a new effective characteristic strain. This iterative process continues until the effective characteristic strain reaches convergence. 

As a result, the shift in characteristic strain, represented by the purple-colored region in \Fig{charac_strain}, becomes reduced compared to the non-eliminated scenario. Furthermore, the frequency band influenced by the SGWB shifts from several $10^{-2}$ Hz down to a few $10^{-3}$ Hz, moving away from LISA's most sensitive range.

Given that the MRs of LISA set the requirements for a minimum SNR level for specific detectable sources, we will further discuss the impact of the SGWB on LISA's detection in terms of variations in SNR in the subsequent sections.

\section{Detectability of MBHBs} \label{MBHBs} 

Massive black hole binaries (MBHBs) are categorized by two types, intermediate mass black holes binaries (IMBHBs) with intermediate mass range between few hundreds and $10^5 \Msun$ for each black holes, and Supermassive black holes binaries (SMBHBs) with mass above $10^5 \Msun$. Tracing the origin, growth and merger history MBHs across cosmic ages is a vital science objectives of LISA. 

The origin of MBHs lurking at the centres of galaxies as power source of active galactic nuclei, is an on-going topic. Some studies predict masses range of the their seeds is around $10^3\Msun$ to fews $10^5 \Msun$, within formation redshift $10\sim15$ \cite{Volonteri:2010wz}. After accretion episodes and repeated merging in the period of cosmic structures clustering, those seeds can grows up to $10^8\Msun$ and more \cite{Sesana:2004sp}. During the growth of seeds, accretion and mergers imprint different information on the spins of them. In order to measure the dimensionless spins and misalignment of spins with the orbital angular momentum with low absolute error, the accumulated SNR (from inspiral phase up to merger) required to reach certain level.

In studying the growth mechanism of MBHs from epoch of the earliest quasars, an accumulated SNR of at least $\sim$ 200 is required to ensure the accurate measurement of parameters. This SNR requirement is also needed for testing the propagation of GWs in LISA’s science investigations (SIs) \cite{LISA:2017pwj}.

In the absence of a SGWB, the expected minimum observation rate of several MBHBs per year would fulfill the requirements of SO2, assuming a conservative population model \cite{Klein:2015hvg}.

\begin{figure}[htbp]
	\centering
	\includegraphics[width=0.48\textwidth]{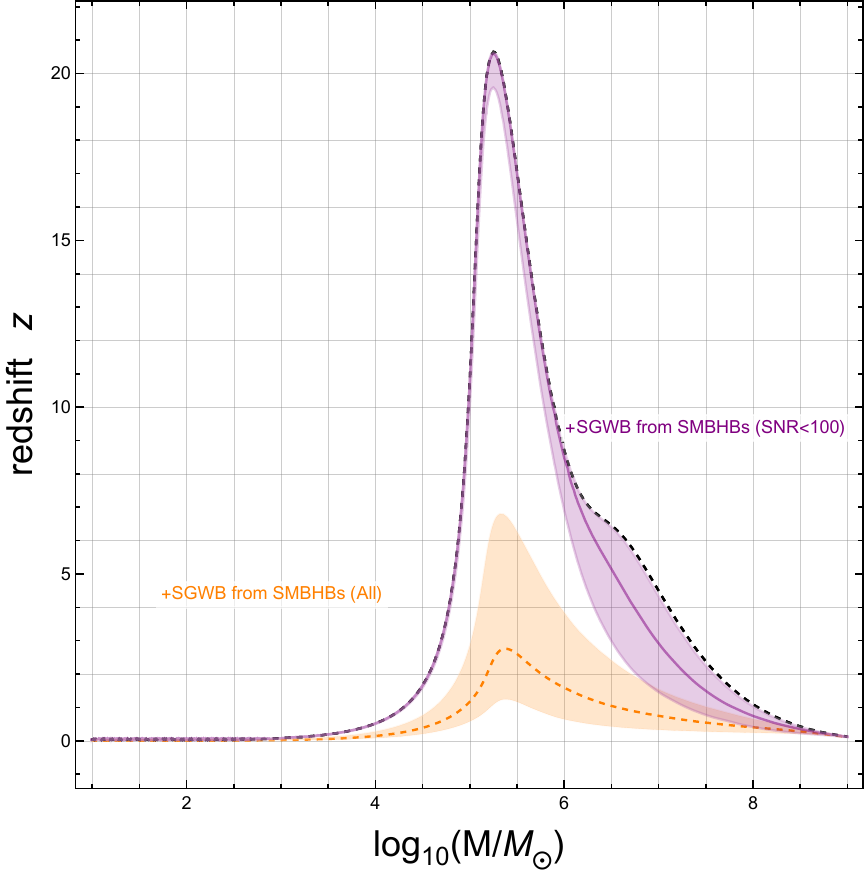}
	\caption{\label{effect on SNR=200}LISA's SNR=200 curves for the GW signals of MBHBs with and without present of SGWB generated by SMBHBs, in the plane of total source-frame with mass $M$, redshift $z$. And the mass ration of binaries  $q$ is 0.2. The black dashed line is curve without  SGWB for SMBHBs. The orange line shows the curve under effect of SGWB from all SMBHBs combined, and the purple line gives the curve under effect of SGWB from SMBHBs that all the signals with SNR$\ge100$ are resolved and eliminated. The shaded region indicates the corresponding $90\%$ credible interval respectively.}
\end{figure}

In \Fig{effect on SNR=200}, we present contour plots of SNR with a value of $\rm{SNR}=200$, using the waveform model derived from references \cite{Ajith:2007kx, Zhu:2011bd}. These plots depict the SNR values for GW signals emitted by MBHBs, both in the presence and absence of the SGWB generated by SMBHBs. The plots are presented in the plane with the source-frame total mass ($M$) and redshift ($z$) of the sources. Without loss of generality, we assume a mass ratio of 0.2 for the binary systems. This choice corresponds to the parameterization used in LISA's Sensitivity Curve SI 2.1, which is designed for the search for seed black holes at cosmic dawn \cite{LISA:2017pwj}.

Our analysis reveals a significant reduction in the detectable redshift of GW signals generated by MBHBs with an SNR of 200 in the mass range $10^4 \sim 10^8\Msun$. This reduction occurs when the SGWB from SMBHBs cannot be resolved and eliminated during the observation period. Given that the goals of LISA SI 2.2 involve the detection of sources at redshift $z<3$ with masses ranging from $10^5$ to $10^6 \Msun$ and an accumulated SNR of at least approximately 200, the presence of the SGWB has a substantial impact on this investigation. However, should we succeed in resolving and eliminating GW signals with SNR values greater than or equal to 100, the impact on the detection of MBHBs in the LISA mission will be significantly reduced. Consequently, the objectives of SI 2.2 will be slightly effected for MBHBs with masses exceeding $10^6$ $\Msun$.

\section{Detectabilities of VGBs and EMRIs}  \label{VGBandEMRI}

Since there are large numbers of compact binaries in the Milky Way galaxy that emit continuous and nearly monochromatic electromagnetic (EM) signals, parts of those binaries are already verified by the observation other than GW detection. For those VGBs emitting GW signal in LISA's frequency band, the joint EM and GW observation can be performed. The details of those VGBs in LISA's band can be found in \cite{Kupfer:2023nqx,VGBdata}. In LISA's SO1: Study the formation and evolution of compact binary stars in the Milk Way Galaxy, the capability to detected and measure the (intrinsic and orbital) parameters of those VGBs is vital. Assuming the strain sensitivity without effect from SGWB, together with estimation of population of VGBs given in \cite{Toonen:2012jj}, LISA should be able to detect and resolve $\sim 25000$ VGBs.

We utilize data from VGBs obtained from Gaia DR3 \citep{Kupfer:2023nqx}. Following the procedures outlined in \citep{Kupfer:2018jee}, the characteristic strain of the gravitational wave (GW) signal emanating from VGBs, as illustrated in \Fig{charac_strain}, is described by
\m
h_c = \sqrt{T_{\rm obs}} \frac{2 (G\mc)^{5/3}}{c^4 d} \pi^{2/3} f^{7/6} .
\n
Here, $\mc$ represents the chirp mass of each individual VGB, $d$ signifies the distance to the source, and $f$ corresponds to the gravitational wave frequency, which is twice the orbital frequency. For the sake of simplicity, the VGBs can be characterized as monochromatic GW signals with a set of parameters obtained from Gaia DR3. It is important to note that the SNR of VGBs can be readily calculated by taking the ratio of the characteristic strain $h_c$ to the effective characteristic strain of the detector $S_{\text{eff}}(f)$.

\begin{figure}[htbp]
	\centering
	\includegraphics[width=0.5\textwidth]{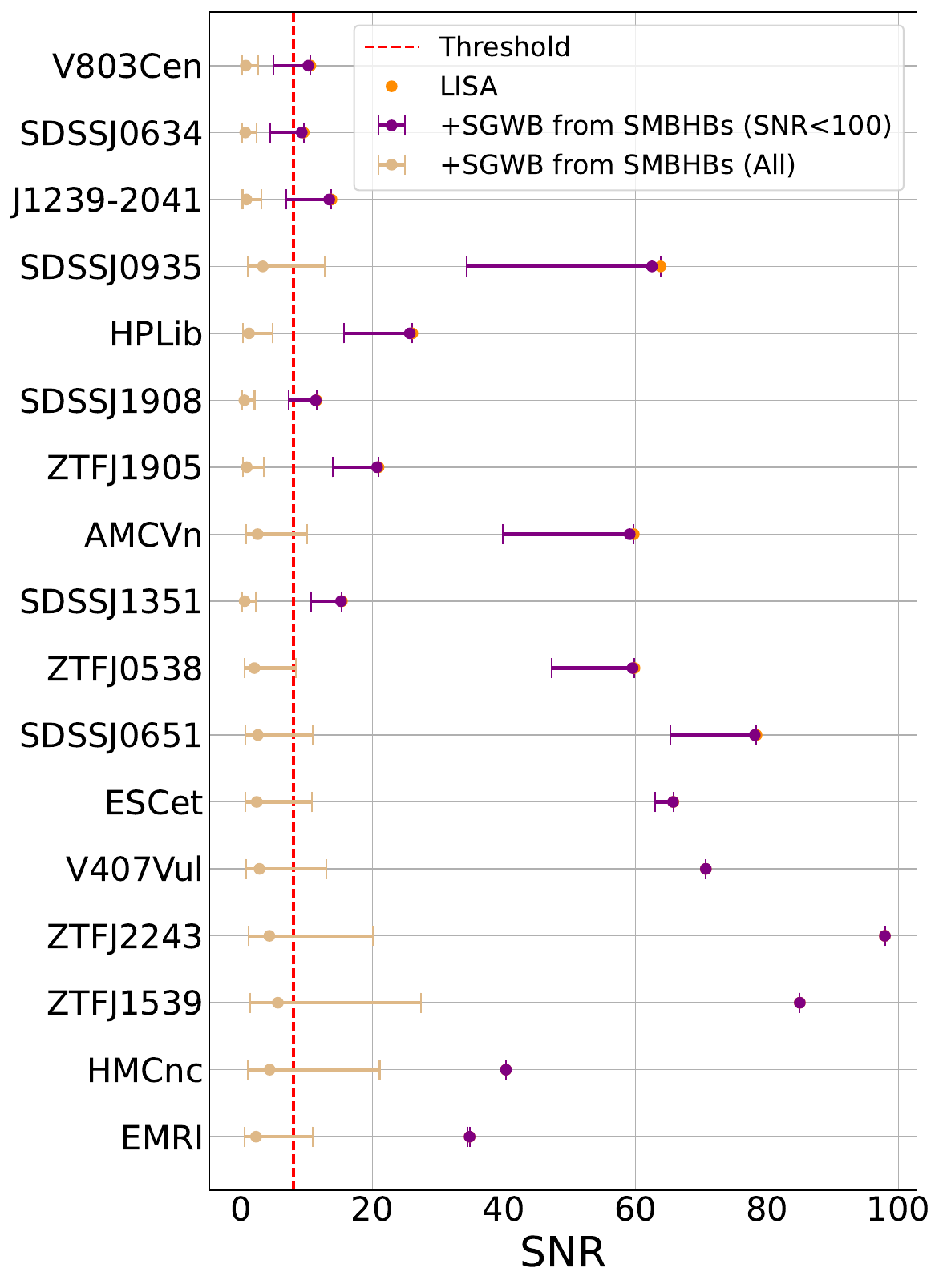}
	\caption{\label{SNR} 
	The SNR for VGB systems and an EMRI example are depicted here. The threshold corresponds to an SNR of 8, signifying the minimum requirement for detecting those compact binary signals. The orange dots represent the SNR for LISA at its original sensitivity level, while the burlywood error bars illustrate the SNR for LISA sensitivity influenced by the SGWB generated by all SMBHBs. The purple error bars points indicate the SNR for LISA's sensitivity effected by SGWB from SMBHBs with SNR values $<100$. For detailed information regarding the names and parameters of VGBs, please refer to \cite{Kupfer:2023nqx, VGBdata}, and for the parameters of the illustrative EMRI signal, kindly refer to Figure \ref{charac_strain}.}
\end{figure}

The Extreme Mass Ratio Inspiral (EMRI) is the inspiral of a stellar-mass compact object, such as a stellar-mass black hole, a neutron star or a white drawf, into a SMBH. Here we adopt the Numerical Kludge method \citep{Babak:2006NK,Chua:2017EMRI} to calculate the time-domain waveform of EMRIs. Because the inspiral signal of EMRIs can stay from months to many years in LISA's frequency band, and the binary might spend up to $10^5$ or more orbits near the innermost stable circular orbit (ISCO), the events of EMRI can give accurate informations of the space-time around the SMBH. Those informations will providing accurate measurements of mass and spin of SMBH, and allowing us to test Kerr geometry in a new high level. In SO3: Probe the dynamics of dense nuclear clusters using EMRIs, and SO5: Explore the fundamental nature of gravity and black holes, the detection of EMRIs in LISA frequency band can open a new channel of astrophysics discovery. Meanwhile, in SO6: Probe the rate of expansion of the Universe, LISA will probe the topics in cosmology via EMRIs.

Because the large uncertainty in the astrophysics of EMRIs, the LISA sensitivity without effect from SGWB would result in a minimum rate of one case per year according to conservative EMRI population models \cite{Amaro-Seoane:2007osp}. For the convenience of presentation, we only take one example of EMRI in \Fig{charac_strain}.

According to the SNR shift caused by the presence of SGWB within LISA noise, as depicted in \Fig{SNR}, it is evident that the SNR of VGBs subjected to the unabated $S_{\rm eff}(f)$ will experience a dramatic reduction. In fact, some of them drop below the threshold level of $\rm{SNR}=8$, rendering them undetectable by LISA. However, the detection of SGWB originating from SMBHBs with $\rm{SNR} \geq 100$ holds the potential to significantly alleviate these SNR shifts for the majority of obscured VGBs, as illustrated in \Fig{SNR}.

\section{Conclusion and Discussion}  \label{Conclusion}

In light of recent discoveries from pulsar timing arrays, we now have compelling evidence of the SGWB produced by SMBHBs for the first time. In this context, we discuss the impacts of the SGWB from SMBHBs on the detection of compact binaries in LISA. It's worth noting that our arguments also apply to other spaceborne GW detectors, such as Taiji \citep{Taiji:2021} and Tianqin \citep{TianQin:2015}. Given that the SGWB from SMBHBs will introduce additional noise, denoted as $S_{\rm GW}(f)$, to LISA's strain sensitivity, it will shift the SNR of MBHBs, EMRIs and VGBs with specific intrinsic and orbital parameters. This influence has the potential to impact the success of the LISA mission.

Based on our analysis, the effective sensitivity will obscure the GW signals of some VGBs and EMRIs, as seen directly in Fig. \ref{charac_strain}, leading to a reduction in their SNR, as depicted in Fig. \ref{SNR}. Furthermore, the detectable redshift under specific SNR conditions for MBHBs will also decrease, as shown in Fig. \ref{effect on SNR=200}. However, if we can resolve and eliminate GW signals from SMBHBs with an SNR above a certain threshold, all these influences can be effectively mitigated.

Therefore, the impacts on the detection of compact binaries due to the SGWB generated by SMBHBs are more significant than the expected galactic background. Understanding the SGWB and investigation their impacts on the LISA mission are crucial for the success of space-borne GW detectors. Additionally, our results may offer an alternative perspective for the design of future GW detectors.

\textit{Acknowledgements.}
We would like to thank Xilong Fan and Lijin Shao for useful conversation. This work makes use of the Black Hole Perturbation Toolkit. This work is supported by the grants from NSFC (Grant No.~12250010, 11975019, 11991052, 12047503), Key Research Program of Frontier Sciences, CAS, Grant No.~ZDBS-LY-7009, CAS Project for Young Scientists in Basic Research YSBR-006, the Key Research Program of the Chinese Academy of Sciences (Grant No.~XDPB15). We acknowledge the use of HPC Cluster of ITP-CAS.

\bibliography{refs}
\end{document}